\documentclass[%
twocolumn,
showpacs,
nofootinbib, nobibnotes,
amsmath,amssymb, aps,
pra,
floatfix,
]{revtex4}
\usepackage{graphicx}
\usepackage{placeins}
\usepackage{dcolumn}
\usepackage{bm}
\usepackage{hyperref}
\usepackage{natbib}
\usepackage{subfigure}

\begin{document}

\preprint{APS/123-QED}

\title{Synchronization of Networked Jahn-Teller Systems in Circuit QED}

%

\author{Yusuf G\"ul}
\email{yusufgul.josephrose@gmail.com.} \affiliation{Faculty of Engineering
and Natural Sciences, Sabanci University, Orhanli-Tuzla, Istanbul,
Turkey} \affiliation{Department of Physics, Bo\u{g}azi\c{c}i
University 34342 Bebek, Istanbul, Turkey}

%
\date{\today}

\begin{abstract}
We consider the nonlinear effects in Jahn-Teller system of two
coupled resonators interacting simultaneously with flux qubit using
Circuit QED. Two frequency description of Jahn Teller system  that
inherits the networked structure of both nonlinear Josephson
Junctions and harmonic oscillators is employed to describe the
synchronous structures in multifrequency scheme. Emergence of
dominating mode is investigated to analyze frequency locking by
eigenvalue spectrum. Rabi Supersplitting is tuned for coupled and uncoupled
synchronous configurations in terms of frequency entrainment
switched by coupling strength between resonators. Second order
coherence functions are employed to investigate self-sustained
oscillations in resonator mode and qubit dephasing. Snychronous
structure between correlations of priviledged mode and qubit is
obtained in localization-delocalization and photon blockade regime
controlled by the population imbalance.
\end{abstract}



\maketitle
\section{Introduction}
Advances in control and flexibility of quantum mechanical
systems leads to the era of quantum simulators such as ultracold
atoms\cite{lewenstein2006,bloch2008}, ion traps
\cite{porras2004,porras2012} and cavity QED
arrays\cite{hartman2008,greentree2006} in the line with the general program of simulating a physical system with another.  Exploring fundamental
quantum mechanics in lattice arrays of Circuit QED
\cite{houck2012,schmidt2013} and  embedding the artificial atoms
into open transmission line resonators
\cite{peropadre2013,abdulmalikov2010} bring another suitability
criteria in strong and ultrastrong coupling regimes for
quantum information processing
\cite{Blais2004,Bourassa2009,Bishop2009,Moon2005}. Emergence of
cooperativity and synchronization in collective behavior of many
body coupled systems trigger quantum classical transitions in small
scales \cite{smerzi1997,raghavan1999,Abbarchi2013,Raftery2104} and
tunable correlations for large scale systems
\cite{Qiu2014,Mendoza2014,Arenas2008,Pikovski2001}.

In solid state systems, Cooperative Jahn-Teller (CJT) systems
 give rise to structural phase
transitions including both linear and quadratic interaction in
definite crystal geometries\cite{popovic2000,kaplan1995}. Previously,$ E
\otimes(b_{1} + b_{2})$ model is investigated for
nonlinear effects by variational principles in  \cite{majernikova2003}.
Nonlinearities lead to the asymmetry of the interaction strengths and the chaotic patterns
in energy levels  of two coupled oscillators
\cite{majernikova2006-1}. Lattice array of localized JT centers are examined as an extension of Dicke model \cite{majernikova2011}.

Superconducting Quantum Interference Devices (SQUIDS) appear as an electrical analog
platform to simulate the collective behavior of particles trapped
in local minima of double well potentials warped by the
nonlinearities \cite{Lefevre-Seguin1992}. Circuit QED architectures inherits the
nonlinear characteristics of Josephson Junctions leading to the
emergence of spatial and temporal transitions in coupled resonator
schemes. Two regimes of coherent oscillations and self trapping by
controlled nonlinearities are analyzed in a hybrid system
\cite{Schoelkopf2008} composed of bosonic Josephson Junction
\cite{Albiez2005,Didier2011}. Localization delocalization
transitions are shown in photon JJ in Circuit QED
setup \cite{Schmidt2010} embedded in a Jaynes-Cummigs (JC) lattice
array \cite{hummer2012,nissen2012}.

In Josephson Junction Arrays (JJAs), due to the collective behavior
of Cooper pairs , synchronization desynchronization transitions
comes out in phase coherence
pattern\cite{Wiesenfeld1996,Wiesenfeld1998,Lin2011}. In spatially
extended systems, cluster of oscillator networks have the ability of
tuning transitions locally as a result of coupling strengths relaxing towards the localized
dominating node\cite{Manzano2013}. Optomechanical systems, as lumped
model of two coupled harmonic oscillator via light, appear as test
bed for reduced form of the effective Kuramato model in dissipative
environment and reconfigurable synchronous oscillator networks
\cite{Heinrich2011,Zhang2012}.

Our aim is to employ Cavity/Circuit QED realization of JT models
\cite{larson2008,Dereli2012} to exhibit the effect of
nonlinearities in multifrequency coupled resonator schemes. Coupled
modes of resonators over which the JT coupling distribution can be
tuned  to manipulate synchronization of qubit dephasing and population
imbalance in terms of normal modes conveying Josephson Junction nonlinearities
\cite{Lecocq2011,Lecocq2012}. Quadratic interactions, responsible
for warping in JT systems networked to the outer circuitry, appear
as the nonlinear Josephson inductance coupling between the flux
qubit and the plasma mode
\cite{Bertet2005-1,Bertet2005-2,Reuther2011}. We present the effect of quadratic interaction in
synchronization of two frequency JT systems. Our system is composed of two coupled resonators interacting simulataneously with a single flux qubit resembling the minimal coupled models of strongly correlated spin-boson systems on a lattice \cite{Ivanov2015, Kurcz2014}.

This paper is organized as follows. In Sec.II we introduce the
coupled model with quadratic interactions and use effective single
mode transformation. The results and discussions are presented in
Sec.III. Finally, we give conclusions in Sec. VI.

\section{Model} \label{sec:model}

Circuit QED simulations of JT-models requires both multi-frequency
description of vibrational interactions and going beyond the
Rotating Wave Approximation (RWA) due to the ultrastrong coupling
regime. Our model hamiltonian is $(\hbar=1)$
\begin{eqnarray}\label{eq:two-resonator cirQED}
H&=&\frac{\omega}{2}\sigma_z+\sum_{i=1,2}\omega_{i}
a^{\dag}_{i}a_{i}+\lambda_{i}(a_{i}+a_{i}^{\dag})\sigma_{x}\nonumber\\
&+&\sum_{i=1,2} g_{i}(a_{i}+a^{\dag}_{i})^{2}\sigma_{x}
\end{eqnarray}
\begin{eqnarray}
H_{NL}&=&[\omega_{eff}(\alpha_{1}+\alpha^{\dag}_{1})^{2}+\omega'(\alpha_{2}+\alpha^{\dag}_{2})^{2}\nonumber\\
&+&J(\alpha_{1}+\alpha^{\dag}_{1})(\alpha_{2}+\alpha^{\dag}_{2})]\sigma_{x}
\end{eqnarray}
where
 $\omega$
and $\omega_{1,2}$ are the qubit and resonator frequencies.
$a_{1,2}(a^{\dag}_{1,2})$ represents the annihilation and creation
operators of resonators and $\sigma_{x},\sigma_{z}$ are the Pauli operators. This hamiltonian
shows the coupling between flux qubit and two plasma mode in both
linear and nonlinear interaction strengths $\lambda_{1,2}$ and
$g_{1,2}$ respectively. To go beyond RWA, Circuit QED realization of
our system is mapped to two frequency JT model and described as
\begin{eqnarray}\label{eq:two-resonator cirQED}
H=H_q+H_{r}+H_{JT}+H_{NL},
\end{eqnarray}
where
\begin{eqnarray}\label{eq:two-resonator cirQED}
H_q=\frac{\omega}{2}\sigma_z
\end{eqnarray}
\begin{eqnarray}\label{eq:two-resonator cirQED}
H_r=\omega_{1}a^{\dag}_{1}a_{1}+\omega_{2}a^{\dag}_{2}a_{2}
\end{eqnarray}
are the qubit and resonator hamiltonians with natural frequencies in
uncoupled scheme. The Jahn-Teller interaction is given by
\begin{eqnarray}\label{eq:two-resonator cirQED}
H_{JT}=k_{1}\omega_{1}( a_{1}+a^{\dag}_{1})\sigma_{x}+k_{2} \omega_{2}(
a_{2}+a^{\dag}_{2})\sigma_{x}
\end{eqnarray} where  $k_{1,2}$
are the dimensionless JT scaling factors
\cite{Dereli2012,O'Brien1972,O'Brien1983}. In the absence of
Nonlinear term, our system behaves as an effective single mode model
where the qubit coupled to resonators asymmetrically due to the
concentration of JT interaction in priviledged mode.

Quadratic interaction terms appear due to the nonlinear Josephson
inductance in SQUID phase leading to the occurrence of second order
terms corresponding to the fluctuations of dynamical variables
controlled by the external parameters
\cite{Bertet2005-1,Reuther2011,Bertet2005-2}.In JT
systems, quadratic interactions are determined empirically  and
depend on the symmetry lowering configurations of crystal
geometries. Then, hamiltonian describing quadratic interactions is written as
\begin{eqnarray}\label{eq:two-resonator cirQED}
H_{NL}=[\omega_{1}( a_{1}+a^{\dag}_{1})^2+ \omega_{2}(
a_{2}+a^{\dag}_{2})^2]\sigma_{x}
\end{eqnarray}
which makes the system networked to outer  crystal structure described  as
bath of harmonic oscillators with natural frequencies
$\omega_{1,2}=g_{1,2}$ in spin-boson treatment. Using both linear
and nonlinear coupling, our system in two frequency effective JT
model\cite{Dereli2012,O'Brien1972,O'Brien1983} becomes
\begin{eqnarray}
H&=& H_{JT}+H_{NL}
\end{eqnarray}
where
\begin{eqnarray}
H_{JT}&=&\frac{\omega}{2}\sigma_z+\omega'\alpha^{\dag}_{2}\alpha_{2}
+J(\alpha^{\dag}_{1}\alpha_{2}+\alpha^{\dag}_{2}\alpha_{1})\nonumber\\
&+&\omega_{eff}[\alpha^{\dag}_{1}\alpha_{1}
+k_{eff}(\alpha_{1}+\alpha^{\dag}_{1})\sigma_{z}]\nonumber\\
&+&c_{2}[(\alpha^{\dag}_{1}\alpha_{2}+\alpha_{1}\alpha^{\dag}_{2})+k_{eff}(\alpha^{\dag}_{2}+\alpha_{2})\sigma_{z}]
\end{eqnarray}
and
\begin{eqnarray}
H_{NL}&=&[\omega_{eff}(\alpha_{1}+\alpha^{\dag}_{1})^{2}+\omega'(\alpha_{2}+\alpha^{\dag}_{2})^{2}\nonumber\\
&+&J(\alpha_{1}+\alpha^{\dag}_{1})(\alpha_{2}+\alpha^{\dag}_{2})]\sigma_{x}
\end{eqnarray}
with the frequency of effective mode
\begin{eqnarray}
\omega_{eff}&=&\frac{\omega_{1}k^{2}_{1}+\omega_{2}k^{2}_{2}}{k_{eff}}
\end{eqnarray}and qubit-resonator coupling strength
\begin{eqnarray}
k^{2}_{eff}&=&k^{2}_{1}+k^{2}_{2}.
\end{eqnarray}
The frequency of the disadvataged mode is given by
\begin{eqnarray}
\omega'&=&\frac{\omega_{1}k^{2}_{2}+\omega_{2}k^{2}_{1}}{k_{eff}}
\end{eqnarray}
and it is coupled to the priviledged mode with a strength
\begin{eqnarray}
c_{2}&=&\frac{\Delta k_{1}k_{2}}{k^{2}_{eff}}.
\end{eqnarray}
where the frequency mismatch $\Delta=\omega_{1}-\omega_{2}$ is used to control
the perturbative interactions on the effective single-mode model.


For simulation purposes, we use two parameters ($k,\Delta$) to see the effect of JT scaling
factors and  frequency
difference of the resonators in going beyond RWA. For this purpose our Circuit QED Hamiltonian is
written as
\begin{eqnarray} \label{eq:cirQED model}
H&=&\hat\alpha_1^\dag\hat\alpha_1+\hat\alpha_2^\dag\hat\alpha_2+\frac{1}{2}\sigma_z+\frac{\Delta}{2}(\hat\alpha_1^\dag\hat\alpha_2
+\hat\alpha_2^\dag\hat\alpha_1)\nonumber\\
&+&\sqrt{2}k[(\hat\alpha_1^\dag+\hat\alpha_1)+(\hat\alpha_1^\dag+\hat\alpha_1)^{2}\nonumber\\
&+&\frac{\Delta}{2}((\hat\alpha_2^\dag+\hat\alpha_2)
+(\hat\alpha_2^\dag+\hat\alpha_2)^{2})]\sigma_x.
\end{eqnarray}

where $k_1=k_2=k$, $\lambda_1=(\omega_1+\omega_2)k/\sqrt{2}$,
$\lambda_2=\Delta k/\sqrt{2}$, and $c_2=\Delta/2$. We present the
coupling of two resonator with the hopping parameter $J=c_{2}.$

In this manner, we consider our coupled system as coupling of
priviledged mode interacting simultaneously with the qubit and the
disadvantaged mode. Correlations of priviledged mode and population imbalance between resonators give rise to cooperative
and synchronous JT systems in Circuit QED.

\section{Results} \label{sec:results}
Externally controlled nonlinearities wired with the JT models
make the coupled systems plausible for emergence of
cooperativity and synchronization in singled out mode
in both strong and ultrastrong regime.

In JJAs, distribution of frequencies are
modulated so as to make the nonlinear oscillators
frequency locked corresponding to the
Kuramoto model of mean field theories \cite{Wiesenfeld1998}. In two
frequency JT model, appearance of singled out effective mode
is investigated in terms of scaling factors $k$ dominating priviledged mode
and frequency difference $\Delta$ representing the coupling strength
of perturbations.

\begin{figure}[h]
\begin{center}
\subfigure[\hspace{0.001cm}]{\label{fig:1a}
\includegraphics[width=0.4\textwidth]{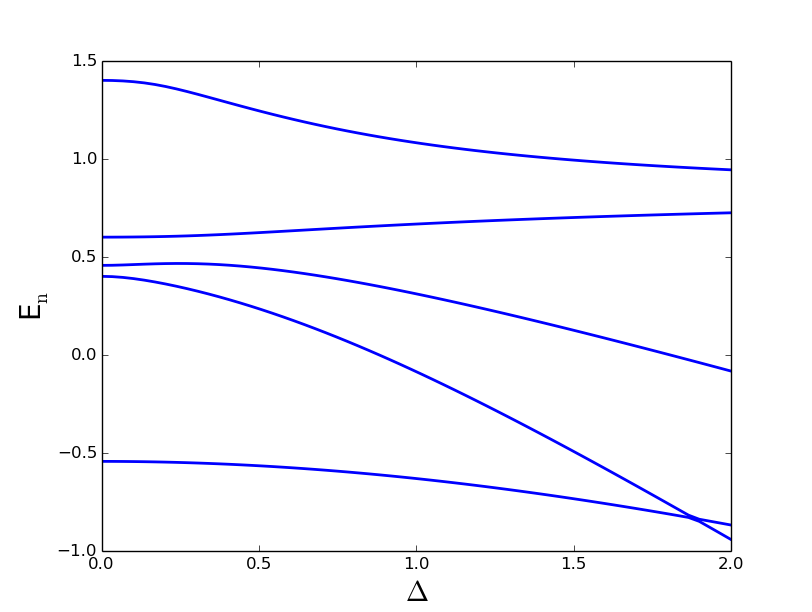}}
\subfigure[\hspace{0.001cm}]{\label{fig:1b}
\includegraphics[width=0.4\textwidth]{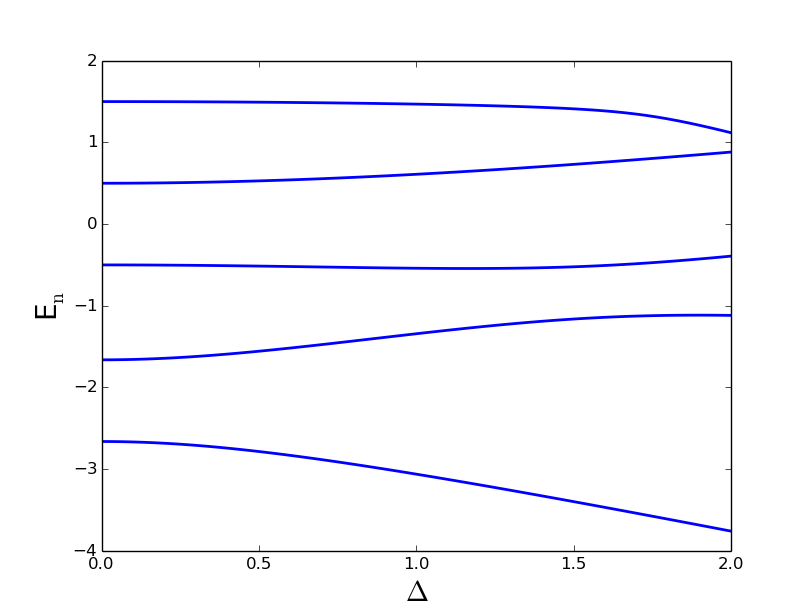}}
\caption{\label{fig3} (Color online)  Emergence of frequency locking
for two-mode JT system shown in spectrum of the lowest five
eigenvalues depending on the frequency difference $\Delta$. (a) At
$\Delta=0$ Rabi splitting of first energy levels occurs for
$k=0.1/\sqrt{2}$. Interaction between priviledged and
disadvantaged mode can be tuned up to $\Delta=0.1$ in single
effective mode.  (b) Range of single mode regime extends up to
$\Delta=0.5$ in ultrastrong regime $k=1.0/\sqrt{2}$ }
\end{center}
\end{figure}

We examined  $5$ lowest eigenenergies in the spectrum of our
system where each resonator is described with Fock space dimension
$2$. In Fig.$2$., wee present the tendency of frequency
locking structure of our system in both strong and ultrastrong
regimes. Our system is in single priviledged mode only
for frequency difference $|\Delta|<0.1$ and $k=0.1/\sqrt{2}$ plotted in fig.$2(a)$. Effect
of perturbative coupling leads to pure Rabi splitting of first
excited level for $|\Delta|=0$. In Fig.$2(b)$, when we are
in ultrastrong regime for $k=1.0/\sqrt{2}$ the range of single mode
structure extends up to $|\Delta|<0.5$ and avoiding crossing is
replaced with a level repelling. The pattern of eigenvalue spectrum is
mixed by $|\Delta|$ and smoothed by $k$ exhibiting the
competition between linear and nonlinear interaction terms.

Circuit QED realizations of vacuum Rabi splitting is detected by the
transmitted amplitude of field quadratures in an array of transmon
 qubits coupled with a common
resonator \cite{Bishop2009}. Linear JT model of two mode
coupled systems shows frequency conversion modulated by nonlinear
susceptibility \cite{Moon2005}. Circuit QED setup is chosen so as to
make it appropriate for transmission measurement and
macroscopic quantum coherence in quantum
classical transition\cite{Fink2010,Fedorov2011}.
\begin{figure}[!hbt]
\begin{center}
\subfigure[\hspace{0.015cm}]{\label{fig:2a}
\includegraphics[width=0.4\textwidth]{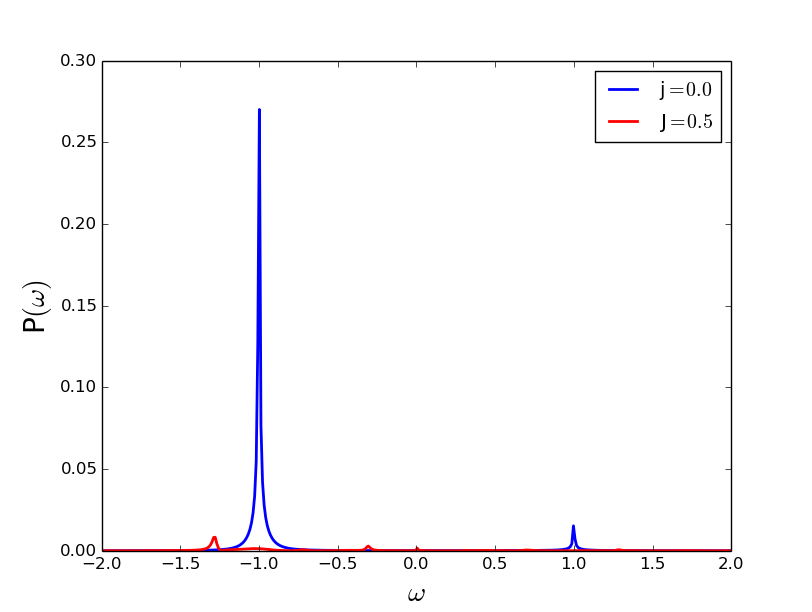}}
\subfigure[\hspace{0.015cm}]{\label{fig:2b}
\includegraphics[width=0.4\textwidth]{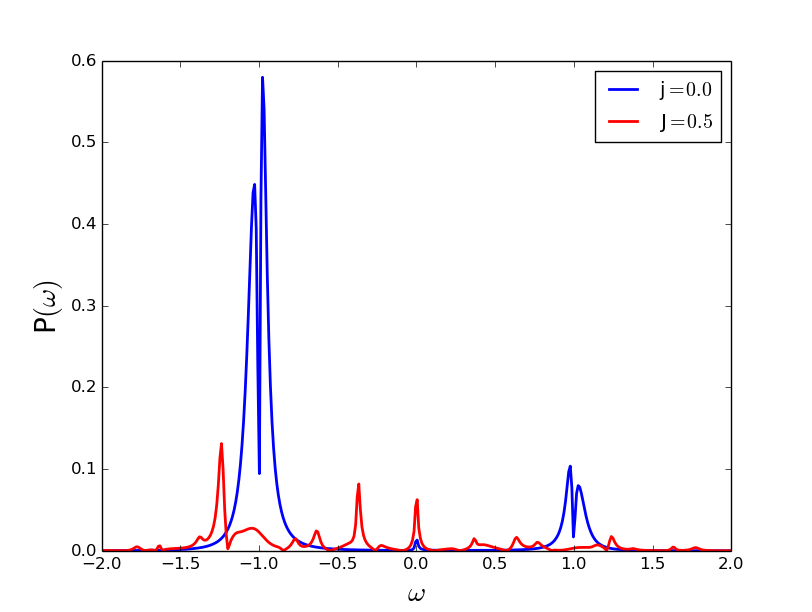}}
\caption{\label{fig3} (Color online) Correlation spectrum of
priviledged mode $\alpha_{1}$ in coupled, and uncoupled scheme
depending on hopping parameter $J$. (a) Asymmetric peaks of
Lorentzian line shape in uncoupled ,$J=0$ and coupled ,$J=0.5$,
resonators in single mode regime $k=0.05\sqrt{2}$ (b) Rabi
supersplitting occurs for $k=0.05$ together with the multilevel
transitions.}
\end{center}
\end{figure}

We use the two time correlation spectrum of heterodyne transmission
spectrum for the privileged mode $\hat\alpha_1$, so that
\begin{eqnarray}
P(\omega)=\int_{-\infty}^{\infty}\langle\hat\alpha_1(t)\hat\alpha_1(0)\rangle
e^{-i\omega t}.
\end{eqnarray}
Open system dynamics is governed by
\begin{eqnarray}
\frac{d\rho}{dt}=-i[H,\rho]+{\cal L}\rho,
\end{eqnarray}
where the Liouvillian superoperator ${\cal L}$ is given by
\begin{eqnarray}
 {\cal L}\rho&=&\sum_{j=1,2}(1+n_{th})\kappa{\cal D}[\hat\alpha_j]\rho+n_{th}\kappa{\cal D}[\hat\alpha_j^\dag]\rho\nonumber\\
 &+&\gamma{\cal D}[\sigma]\rho+\frac{\gamma_\phi}{2}{\cal D}[\sigma_z]\rho,
 \end{eqnarray}
with $n_{th}$ representing the  average thermal photon number. Taking 
$n_{th}=0.15$ corresponds to $100$~mK \cite{Fink2010,Dereli2012}.
${\cal D}$ denotes the Lindblad type damping superoperators,
$\kappa$ is the cavity photon loss rate. Qubit relaxation and
dephasing rates are, respectively, $\gamma$ and $\gamma_\phi$.We use
balanced dissipation where resonator decay parameters
$\kappa_{1}=\kappa_{2}=0.001$ and qubit relaxation and dephasing
parameters $\gamma=0.001, \gamma_{\phi}=0.01$ with the thermal
occupation number $n_{th}=0.15$.

Fig.$2$ shows nonlinear vacuum response of the cavity field for
hopping parameter J$=0$ and J$=0.5$ values . Fig.$2(a)$ presents
asymmetric Rabi peaks in Lorentzian line shape when the system is in
single mode JC regime. Increasing the  coupling strength  reveals
the emergence of supersplitting of each vacuum Rabi peak into a
doublet with a higher amplitude. Fig.$2(b)$ shows the  Rabi
supersplitting for $k=0.05$. Going beyond the single mode JC regime
increase the central dip in each peaks. Effect of hopping parameters
is seen in multi-photon transitions.

\begin{figure}[!hbt]
\begin{center}
\subfigure[\hspace{0.015cm}]{\label{fig:3a}
\includegraphics[width=0.4\textwidth]{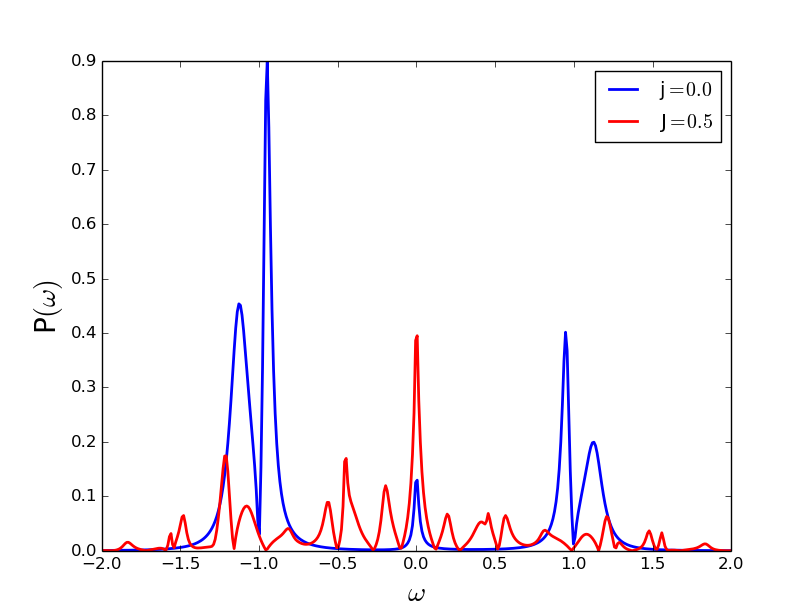}}
\subfigure[\hspace{0.015cm}]{\label{fig:3b}
\includegraphics[width=0.4\textwidth]{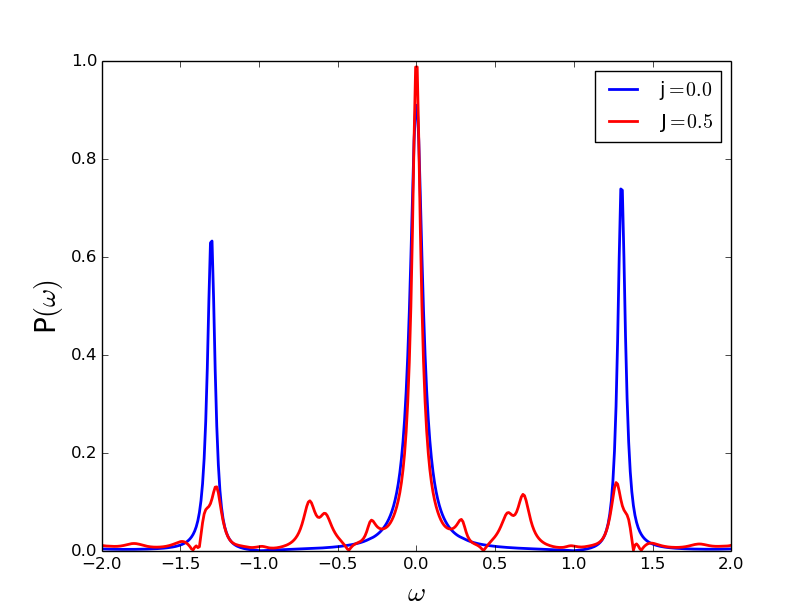}}
\caption{\label{fig3} (Color online)  Appearance of stokes
anti-stokes peaks in two mode JC regime. (a)Shift of side peaks in
strong coupling regime $k=0.1/\sqrt{2}$ and nonlinear response of
the system is shown by the increase central dip of each side peak
(b) At intermediate coupling regime $k=0.5/\sqrt{2}$. asymmetry
occurs in side peaks and amplitude of the central peak gets higher
than the others. Coincidence of the central and side peaks is used
for the synchronous schemes for coupled and uncoupled
configurations. }
\end{center}
\end{figure}

when the number of subpopulations
in variants of Kuramoto model is equal to the degrees of freedom of
the system under considerations, one can obtain reconfigurations of
coupled and uncoupled schemes by tuning frequency mismatch with
$(k,\Delta)$ parameters. In coupled resonator scheme, $\Delta\neq0$,
hopping terms appears as the inter-cavity control parameter
resulting in cooperative and synchronous structure. Field
quadratures are used in describing measure of quantum analogue of
frequency entrainment and locking in optomechanical systems and
harmonically driven Van der Pol
oscillator\cite{Mari2013,Walter2014}. Self-sustained oscillation is
shown in amplitude locking with quadratic coupling leading to
multipeak field spectrum \cite{Zhang2014}.

Fig. $3$ presents how the coupling regime dominates the inherited
features of nonlinear Josephson Junctions depending on the
configurations. Fig. $3(a)$ shows the shift of splitted peaks from
each other and raise of extra peak around $\omega=0$ for $k=0.1$
which indicates the synchronization
entrainment although their amplitude is still different. In
Fig.$(3b)$, at intermediate coupling regime $k=0.5$, spectrum evolve
into a triplet where the asymmetry of peaks are tuned in terms of relative
coupling between priviledged and disadvantaged mode. Emergence of
stokes and anti-stokes peaks are due to field quadrature operator which
behaves as qubit-polariton operator revealing Raman process.
Coherent evolution is modulated by the multilevel structure of
atomic states carrying the nonlinearities of JJs intrinsically.
The frequency amplitude of coupled and
uncoupled scheme gets closer and is coincident in both central and
side peaks in definite frequencies. Switchable synchronous configurations are obtained by synchronization entrainment between the pure two mode
JT  and the effective priviledged mode model.

Two frequency realization of Circuit QED architectures appears as a
platform to simulate the self-sustained oscillator behavior of
tedrahedral networks distorted by corner sharing spin, where
nonlinearities are induced by the lattice restoring energy.
 Switching symmetric and asymmetric mode
configurations leads to the transverse and longitudinal prolongation
of host lattice arrangement mimicking the rhythmic behavior of
diamond shaped crystal geometries. Accumulate and fire oscillators
description give way to slow growth of correlations similar to the van der Pol relaxation oscillator
\cite{Mari2013,Walter2014,Pikovski2001} in the presence of
distortions. In our model, self-sustained oscillation of each normal
mode is described by correlations revealing delocalization and
trapping regimes of coupled system.

In order to see the correlations of distortions, we use the second
order coherence functions of field and atomic states
\begin{eqnarray}
g_{i}^{(2)}=\frac{O^{\dag}_{i}(t)O^{\dag}_{i}(t+\tau)O_{i}(t)O_{i}(t+\tau)}{O_{i}^{\dag}(t)O_{i}(t)}
\end{eqnarray}
where $i=r,q$ are used in place of the resonator and the qubit respectively.

The condition $g^{(2)}_{r,q}\ll1$ corresponds to antibunching, and
used for the indication of photon blockade and  energetic
localization of qubit. Another central quantity of coupled cavity
system is the photon population imbalance
$z(t)=(n_{1}-n_{2})/(n_{1}+n_{2})$ where $n_{i}=$Tr$
\hat{\alpha}_{i}^{\dag}\hat{\alpha}_{i}\hat{\rho}$ for $i=1,2$
corresponds to the two cavities described by priviledged and
disadvantaged mode. Total photon number is given by $N=n_{1}+n_{2}$.
In performing calculations we take $N=5$ and both of the priviledged
mode and qubit are excited initially.

\begin{figure}[!hbt]
\begin{center}
\subfigure[\hspace{0.010cm}]{\label{fig:4a}
\includegraphics[width=0.5\textwidth]{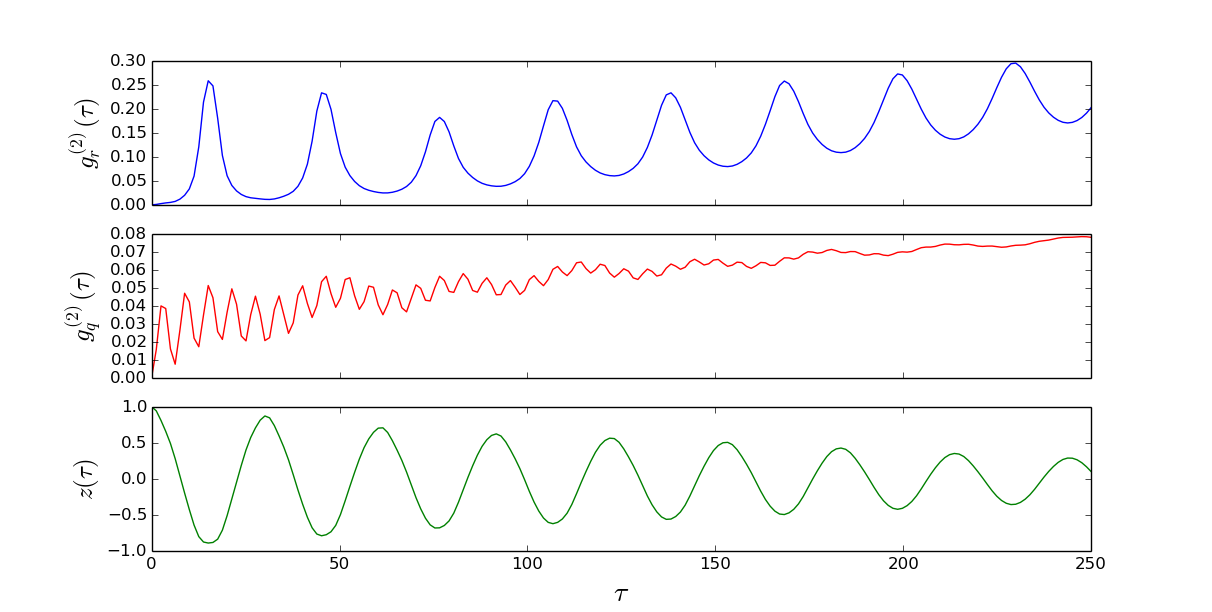}}
\subfigure[\hspace{0.010cm}]{\label{fig:4b}
\includegraphics[width=0.5\textwidth]{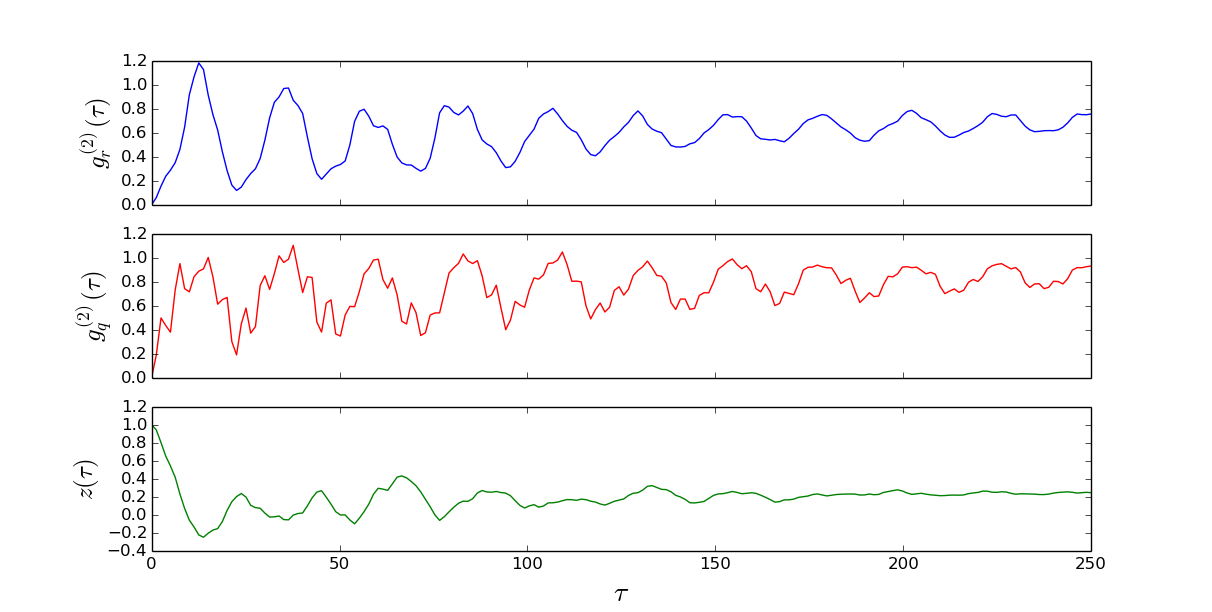}}
\subfigure[\hspace{0.010cm}]{\label{fig:4c}
\includegraphics[width=0.5\textwidth]{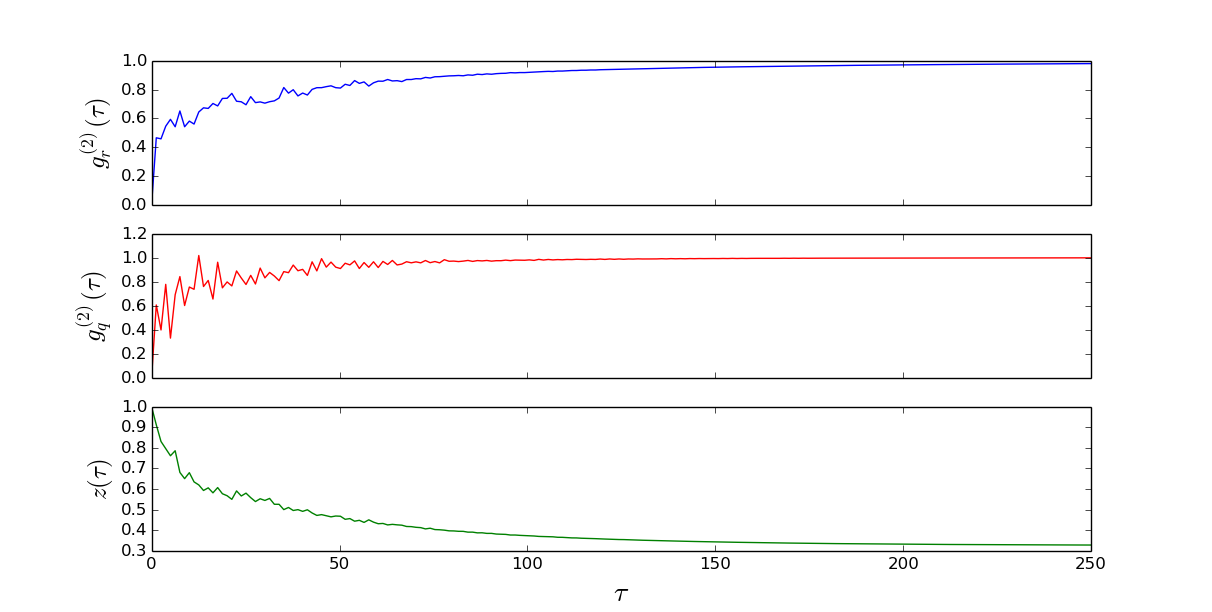}}
\caption{\label{fig3} (Color online) Emergence of localization and
synchronization transitions in weak, strong and ultrastrong regime.
(a) shows snynchronous structure between damped oscillating
population imbalance and correlation of priviledged mode in weak
coupling $k=0.01/\sqrt{2}$. (b) priviledged mode becomes synchronous
with qubit and delocalization-localization transition occurs in
population imbalance in strong coupling regime $k=0.1/\sqrt{2}$. (c)
presents the photon blockade in priviledged mode and fully trapped
regime in population imbalance with $k=1.0/\sqrt{2}$ and
$\gamma_{\phi}=0.1$ }
\end{center}
\end{figure}

In Fig.$4$ we show the correlations functions of resonator and qubit
in weak, strong and ultrastrong regimes. In the first two top panels,
we present the second order coherence functions of priviledged mode
(blue) and the qubit (red) and the third panel shows population
imbalance (green). Fig.$4.a$ shows correlations and poulation
imbalance in weak coupling regime, $k=0.01/\sqrt{2}$. Population
imbalance is in oscillating regime and synchronous with the photon
correlation. Self-sustained oscillation is seen via decreasing of
population imbalance while resetting of antibunching of priviledged
mode at two different time scales corresponding to accumulate and
fire oscillator in the sense of van der Pol relaxation.  As $\tau$
increases population imbalance reaches zero and qubit correlation
with beats in anharmonic time intervals become stable. Starting
with $g^{(2)}_{r,q}(0)=0$ corresponding to the photon blockade
regime, photon correlations reach stable point with decreasing peaks
while onsite repulsion is increasing. In strong regime
$k=0.1/\sqrt{2}$, fig.$4$.b shows the delocalization-localization
transitions in population imbalance. Contrary to the weak coupling
regime, qubit and photon correlation becomes synchronous
representing simultaneous firing and damping of correlations as
$\tau$ increases. Although accumulation gets diminished in blockade
regime, there is still firing of qubit and priviledged mode
correlation due to the multilevel tansitions by the inherited
nonlinearities of JJs. Fig.$4.$c presents the fully localized regime
for population imbalance and all the quantities reach a stable point
in ultra-strong regime,$k=1.0/\sqrt{2}$. Effect of qubit dephasing
is shown by quenching thermal fluctuations by taking
$\gamma_{\phi}=0.1$ and leaving the other parameters the same.

These results suggest that two frequency description JT system can
be used to investigate cooperative and synchronous behaviors of
circuit QED schemes by modulating the qubit anharmonicities due to
JJs with networked nonlinearities.
\section{CONCLUSION}\label{sec:conclusion}
In conclusion, we have shown that nonlinearities play a central
role in describing the cooperativity and synchronization in Circuit
QED architectures by the inherited nonlinearities of Josephson Junctions. In our
model, flux qubit simultaneously is coupled to two resonator with both
linear and quadratic interaction terms. We performed the eigenenergy
and power spectrum calculations in frequency locking and
synchronization entrainment regimes respectively. Nonlinearities
give way to  exploration of quantum mechanics at the fundamental level such as Rabi
Super-splitting. Tedrahedral structures of JT systems opens the way
of constructing networked oscillators which can be translated to
coupled resonator schemes of circuit QED. Correlation functions of
normal modes and qubit indicate cooperative and synchronous
structures in localization delocalization and photon blockade
regimes.

\begin{acknowledgements}

Y. G. gratefully acknowledges support by Bo\u{g}azi\c{c}i University
BAP project no $6942$.
\end{acknowledgements}

%

\end{document}